# Active Phase and Amplitude Modulation in Perovskite Polaritonic Waveguides

Andrea Zacheo[1,2,†], Marco Marangi[1,2,†], Tim Colin Meiler[1,2,†], Filippo Martinelli[1,2],

Antonio Fieramosca[3], Vincenzo Ardizzone[3], Giorgio Adamo[1,2,4], Daniele Sanvitto[1,2,3], Cesare Soci[1,2,5,*]

[1]*Centre for Disruptive Photonic Technologies, The Photonics Institute, Nanyang Technological University, Singapore 637371*

[2]*Division of Physics and Applied Physics, School of Physical and Mathematical Sciences, Nanyang Technological University, Singapore 6373713*

[3]*CNR NANOTEC Institute of Nanotechnology, Via Monteroni, Lecce 73100, Italy*

[4]*Optoelectronics Research Centre, University of Southampton, Highfield, Southampton SO17 1BJ, UK*

[5]*School of Electrical and Electronic Engineering, Nanyang Technological University, Singapore 639798, Singapore*

[†] *These authors contributed equally to this work*

* *Email: csoci@ntu.edu.sg*

**Integrated photonic technologies require active materials capable of providing large optical modulation within compact footprints and low losses. The large optical nonlinearities of exciton-polaritons offer a promising route to meet these requirements, yet their integration into photonic circuits remains challenging because strong light-matter interactions must be combined with low-loss transport and spatially localized control. Here, we demonstrate active phase and amplitude modulation in halide perovskite polaritonic waveguides integrated with electrical microheaters. Local electrothermal actuation drives a structural phase transition in the perovskite, modifying the complex refractive index dispersion near the excitonic resonance. By tuning the excitonic fraction of the propagating mode, the waveguides can therefore operate either as a phase shifter with a tuning efficiency of $P_\pi \cdot L_\pi \approx 0.2\ \mathrm{mW \cdot mm}$, approximately one order of magnitude lower than silicon modulators with similar geometries, or as an intensity switch , providing $1\ \mathrm{dB/\mu m}$ modulation depth at $20\ \mathrm{mW}$. Our work establishes phase-change excitonic reconfiguration as a route towards compact polariton circuits for nonlinear photonic applications.**

## Introduction

Photonic integrated circuits (PICs) offer a scalable route to transport and process information with high bandwidth and low propagation loss. Extending their functionality beyond passive routing towards programmable optical processing, however, requires active materials and devices capable of controlling the phase and amplitude of light over short interaction lengths.[1] This requirement is particularly stringent for photonic logic operations,[2] quantum information processing,[3] and neuromorphic computing,[4] where large numbers of processing nodes, such as phase shifters and optical switches, must be integrated on a single chip. Conventional thermo-optic phase shifters are scalable and straightforward to implement but typically rely on weak refractive index modulation, resulting in large device footprints and high power consumption.[5,6] Electro-optic approaches offer faster operation, but often require similarly long device lengths because of their modest index modulation, while also introducing additional complexity associated with the integration of active materials.[7,8] Resonant architectures such as microring resonators can achieve compact, low-power electro-optic modulation through resonant enhancement, although at the cost of limited bandwidth and increased sensitivity to resonance drift.[9] Conventional phase-change materials (PCMs) such as $Ge_2Sb_2Te_5$ (GST),[10] $Ge_2Sb_2Se_4Te_1$ (GSST),[11] $VO_2$,[12] and the more recent $Sb_2S_3$,[13] and $Sb_2Se_3$,[14] can instead provide substantially larger optical constant variations within a considerably more compact modulation regions. Their optical band edge is characterized by a broad, highly absorptive continuum, and the structural rearrangement between material phases modifies the electronic polarizability and, consequently, the refractive index.[15] Because both refractive index and absorption originate from the same electronic response, reducing optical losses generally requires operating far below the bandgap, where the achievable index contrast is also reduced. This trade-off restricts the accessible operating energies and has largely confined PCMs photonics to the telecommunication band.[16] This limitation can be overcome by exploiting the steep dispersion associated with excitonic resonances. Unlike a broad absorption continuum, an exciton concentrates its oscillator strength within a narrow spectral range, allowing a large refractive index variation to persist on the dispersive tail where absorption is substantially reduced. Even a modest spectral shift of the exciton can therefore produce a large refractive index change at an operating energy with negligible optical loss. Therefore, an "excitonic PCM", a phase-change material that exploits this mechanism by shifting the excitonic resonance across a structural transition, could extend efficient modulation toward largely unexplored spectral regions, in particular the near-infrared.[17] This spectral region is becoming increasingly relevant for quantum and photonic technologies, encompassing the optical transitions of rubidium atoms used in quantum memories,[18] and quantum sensors,[19] the emission of several semiconductor quantum dots,[20] and the broad gain spectrum of Ti:sapphire.[21] Beyond enabling dispersive low-loss modulation, excitonic resonances can be strongly coupled directly to guided photonic modes to form exciton-polaritons, hybrid light-matter quasiparticles that combine the propagation properties of guided photons,[22] with the strong dispersion and nonlinear interactions of excitons.[23] The excitonic fraction of these particles provides a tunable parameter through which the sensitivity of the propagating mode can be continuously controlled. Guided polaritons are particularly attractive because, unlike microcavity polaritons whose photonic component is primarily confined along the vertical direction,[24] they are intrinsically propagating modes confined by total internal reflection within a semiconductor slab, thereby directly integrating polaritonic nonlinearities into planar photonic devices such as waveguides,[25] interferometers,[26] or photonic-crystal slabs.[27] Waveguide polaritons were first demonstrated in GaAs/AlGaAs quantum-well heterostructures,[28] and have enabled the demonstration of nonlinear continuum generation,[29] electrically controlled dipolariton transport and interactions,[30–32] electrically controlled polariton lasing,[33] enhanced nonlinear parametric effects,[34] and topological edge state routing.[35] Moreover, wide-bandgap semiconductors such as ZnO, AlInGaN, and GaN have extended guided polaritonics to room temperature and shorter wavelengths,[36–38] while transition-metal dichalcogenides have enabled ultrathin guided modes with electrostatic tunability.[39] Within this rapidly expanding landscape, halide perovskites have emerged as a particularly attractive material platform for photonics and polaritonics circuits.[40] Their large exciton binding energies and oscillator strengths sustain the strong-coupling regime at elevated temperatures and give rise to strong optical nonlinearities. At the same time, their high refractive index, solution processability, and compatibility with nanofabrication, enable optical confinement and excitonic functionality to coexist within the same material layer.[41] These properties have been extensively exploited in single-crystalline $CsPbBr_3$ for room-temperature microlasers,[42] and small passive photonic circuits,[43] as well as in polaritonic waveguides exhibiting condensation and edge lasing,[44] chiral transport,[45] and optical amplification.[46] Guided strong coupling has also been demonstrated in microstructured two-dimensional Dion-Jacobson perovskites,[47] while large scale polycrystalline $MAPbBr_3$ film waveguides have enabled room-temperature polariton-driven self-phase modulation.[48]

Here, we demonstrate electrical control of propagating exciton-polaritons in integrated perovskite waveguides by exploiting the structural phase transition of polycrystalline methylammonium lead iodide ($MAPbI_3$). Across this transition, $MAPbI_3$ acts as an excitonic PCM, shifting the exciton resonance and locally modifying the complex refractive index. This reconfigures the polariton dispersion and the excitonic fraction, enabling direct control over the phase and amplitude of the propagating mode in the near-infrared region. Spatially selective actuation is achieved using electrically driven microheaters adjacent to the waveguide. At lower excitonic fraction, the device operates as a phase shifter, providing a phase shift of up to π over a modulation length $L_\pi$ of only $10\ \mu\text{m}$, corresponding to a tuning efficiency $P_\pi L_\pi \approx 0.2\ \text{mW} \cdot \text{mm}$, that is approximately one order of magnitude lower than in silicon modulators with comparable geometries.[49] At higher excitonic fraction, the increased overlap with the absorptive excitonic resonance leads to stronger propagation losses, transforming the same device into an intensity switch with $1\ \text{dB}/\mu\text{m}$ modulation at $20\ \text{mW}$. The excitonic fraction therefore acts as a spectral control parameter that determines both magnitude and character of the optical modulation within the same waveguide architecture. These results establish an excitonic route to phase-change photonics in which structural reconfiguration acts through a strongly dispersive material resonance of the excitonic PCM whose optical response is determined by the exciton-photon composition of the propagating polariton mode. This concept could also be extended across the visible and near-infrared spectral regions through compositional engineering of halide perovskites,[50] enabling compact and reconfigurable components for classical and quantum heterogeneous PICs.[51,52]

## On-chip integration of polariton waveguides in $MAPbI_3$ perovskite films

Previous studies have demonstrated strong exciton-photon coupling,[53] polariton condensation,[54] and electrically injected polaritonic states in $MAPbI_3$ thin films.[55] Here, we translate these properties into polaritonic circuits, with a single perovskite layer providing both optical confinement and excitonic functionality. **Figure 1a** schematically illustrates the device platform, comprising patterned waveguides, splitters and grating couplers for launching, routing and collecting propagating polaritons. Injected photons couple into the waveguide through the input grating, hybridizing with the $MAPbI_3$ excitonic resonance to form guided polariton modes, which subsequently propagate through the patterned circuit and are collected at the output gratings. Active control of propagating polaritons is achieved by integrating an electrical microheater adjacent to one arm of the Y splitter. Local Joule heating drives the orthorhombic-to-tetragonal structural transition of the perovskite crystalline lattice that occurs at a critical temperature $T_C \approx 130-135$ K, red-shifting the excitonic resonance of about 60 meV and thereby modifying the polariton dispersion and propagation. At lower energies where the polariton retains a relatively low excitonic fraction, the thermal excitonic reconfiguration primarily produces phase modulation. At higher energies, closer to the unperturbed exciton resonance, the larger excitonic fraction enhances the response to the same electrothermal perturbation and leads instead to a pronounced modulation of the transmitted intensity. The same integrated element can therefore operate as either a phase shifter or an optical switch by selecting the energy of the propagating polariton.

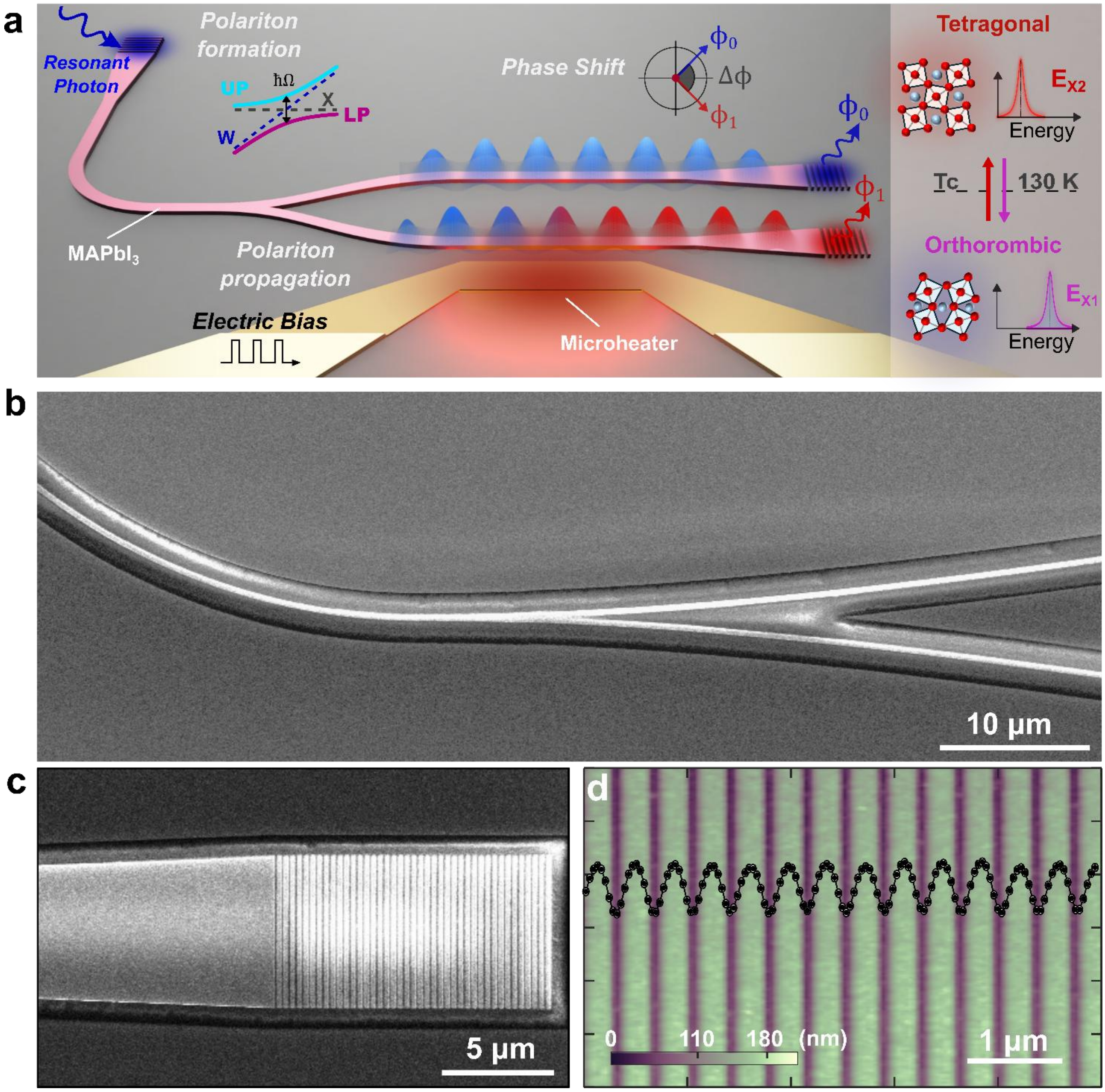

**Figure 1. Polaritonic circuits in $MAPbI_3$ perovskite thin film**: **(a)** Schematic of the polaritonic circuit architecture, comprising grating couplers, waveguides and splitters patterned in a solution-processed $MAPbI_3$ film, that allow generation and propagation of the exciton-polaritons. Electrically driven microheaters integrated alongside selected waveguide sections locally shift the excitonic resonance by inducing the orthorhombic-to-tetragonal structural transition of perovskite, thereby enabling electrothermal control of polariton propagation. **(b)** Scanning electron microscopy image (SEM) of the splitter, defined into the film by focused ion beam milling. **(c,d)** High-magnification SEM image of a diffraction-grating coupler **(c)** and corresponding AFM image with horizontal height profile **(d)**, revealing the shallow periodic modulation designed for weakly perturbative outcoupling. The period of the grating is 385 nm.

To realize this functionality in an integrated platform, $MAPbI_3$ films were deposited by spin coating from solution onto silicon substrates coated with a 3 μm thick $SiO_2$ layer. The films were subsequently patterned by focused ion beam milling in two sequential fabrication steps and finally encapsulated with PMMA (see *Methods*). The scanning electron microscopy image in **Fig. 1b** shows a representative Y splitter with connected input and output branches, while the higher magnification image in **Fig. 1c** illustrates the periodic structure of the grating coupler. Atomic force microscopy further resolves the periodic surface modulation and corresponding height profile (**Fig. 1d**), showing the shallow groove depth selected to provide optical outcoupling while minimizing perturbation of the propagating polariton mode. The waveguide geometry is designed to support only the fundamental transverse-electric polaritonic mode $LP_0$, with a width of 330 nm and a thickness of approximately 140 nm. The single mode configuration suppresses modal interference and maintains strong confinement of the propagating polariton field (**Fig. 2a**). The calculated effective index increases as the guided mode approaches the orthorhombic $MAPbI_3$ exciton resonance at 1.685 eV, while the group velocity progressively decreases, reflecting the stronger dispersive contribution of the excitonic component (**Fig. 2b**). The polariton dispersion was experimentally mapped using non-resonant excitation focused 5-7 μm away from the grating coupler with a period of 385 nm, allowing a broad range of guided states, including those with a larger excitonic fraction, to be populated. The measured dispersion in **Fig. 2c** exhibits a pronounced curvature near the exciton energy and is well reproduced by a coupled-oscillator model, yielding a Rabi splitting of approximately 107 meV. The observation of a single lower-polariton branch is consistent with the simulated single-mode waveguide response. Resonant transmission measurements were subsequently performed at 1.5498 eV (800 nm) as show in in **Fig. 2c**, where the propagating state retains a sufficiently highly photonic character ($|X|^2 \approx 0.2$) to sustain low-loss transport. At this energy, polaritons propagate over distances exceeding 300 μm and are clearly observable at the grating outcoupler after travelling through the waveguide (see *Supplementary Information*). This propagation enables the integration of multiple photonic components within the same polaritonic circuit, including splitters, interferometric sections and locally addressable modulators.

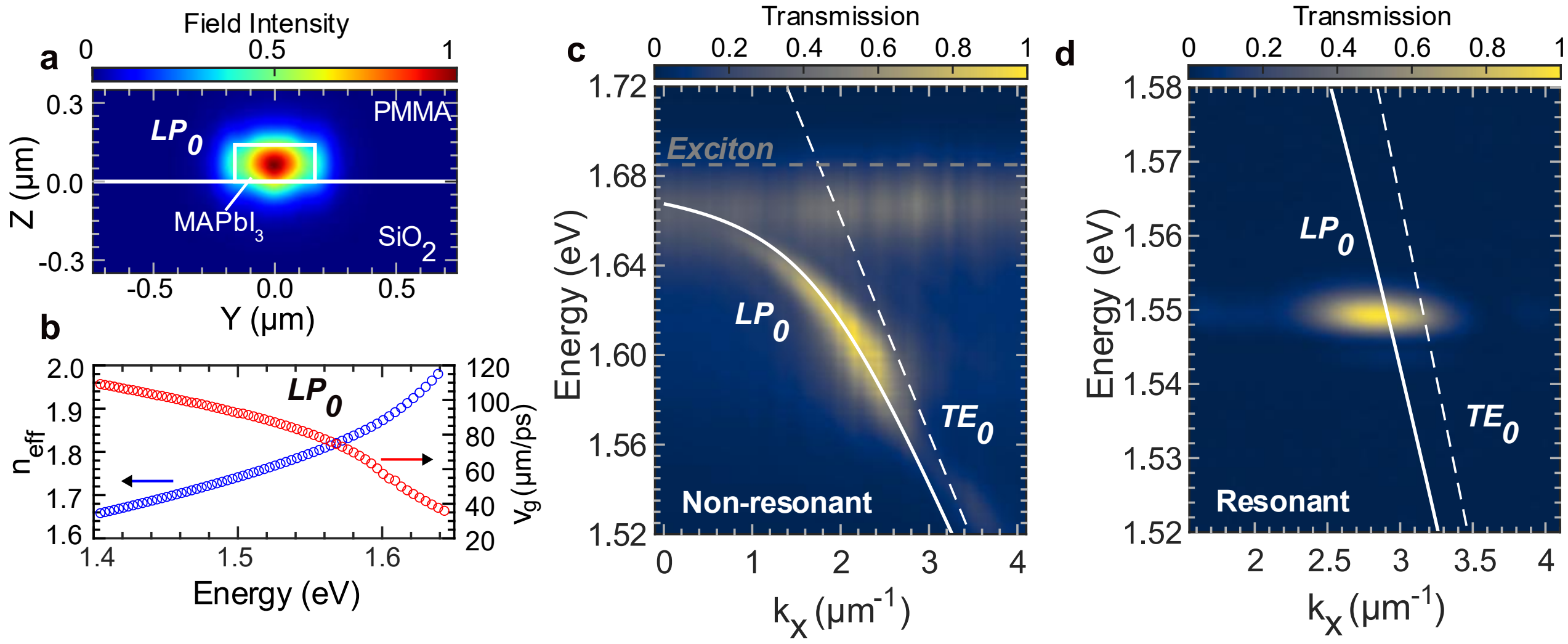

**Figure 2. Long-range propagation of single-mode guided polaritons**: **a)** Simulated electric field intensity distribution of the fundamental mode supported by a 330 nm wide $MAPbI_3$ waveguide. **b)** Calculated effective refractive index $n_{eff}$ and group velocity $v_g$ of the mode as a function of energy, showing increasing dispersion and reduced group velocity toward the excitonic resonance (1.685 eV). **c)** Momentum-resolved transmission map showing the guided polariton dispersion under non-resonant excitation. The solid white line represents the lower-polariton branch obtained from a coupled-oscillator fit, while the white diagonal and grey horizontal dashed lines indicate the uncoupled $TE_0$ mode and $MAPbI_3$ exciton, respectively. **d)** Momentum-resolved transmission collected from the output grating under resonant excitation at 1.550 eV, demonstrating coupling to the guided lower-polariton state. The solid and dashed lines show the calculated lower-polariton and uncoupled photonic dispersions, respectively.

## Phase modulation of propagating polaritons

The active device consists of an input polariton waveguide connected to a Y-splitter with two output arms, one of which is coupled to a resistive chromium microheater for local electrical control (**Fig. 3a**). The 10 µm long microheater is positioned approximately 3 µm from the waveguide, a separation optimized to maintain efficient thermal coupling while preventing direct overlap between the metallic heater and the guided optical mode. This geometry therefore provides a compromise between maximizing the local modulation strength and minimizing additional optical absorption and scattering, while preserving the structural continuity of the perovskite waveguide. To further confine the temperature increase and minimize thermal crosstalk between the two arms, the substrate stack was engineered to control heat diffusion. The 3 μm thick oxide layer reduces heat dissipation into the underlying silicon substrate and promotes lateral thermal confinement. Finite-element simulations show that the temperature field generated by the microheater efficiently reaches the addressed waveguide while maintaining a pronounced thermal asymmetry between the two arms (**Fig. 3b**).

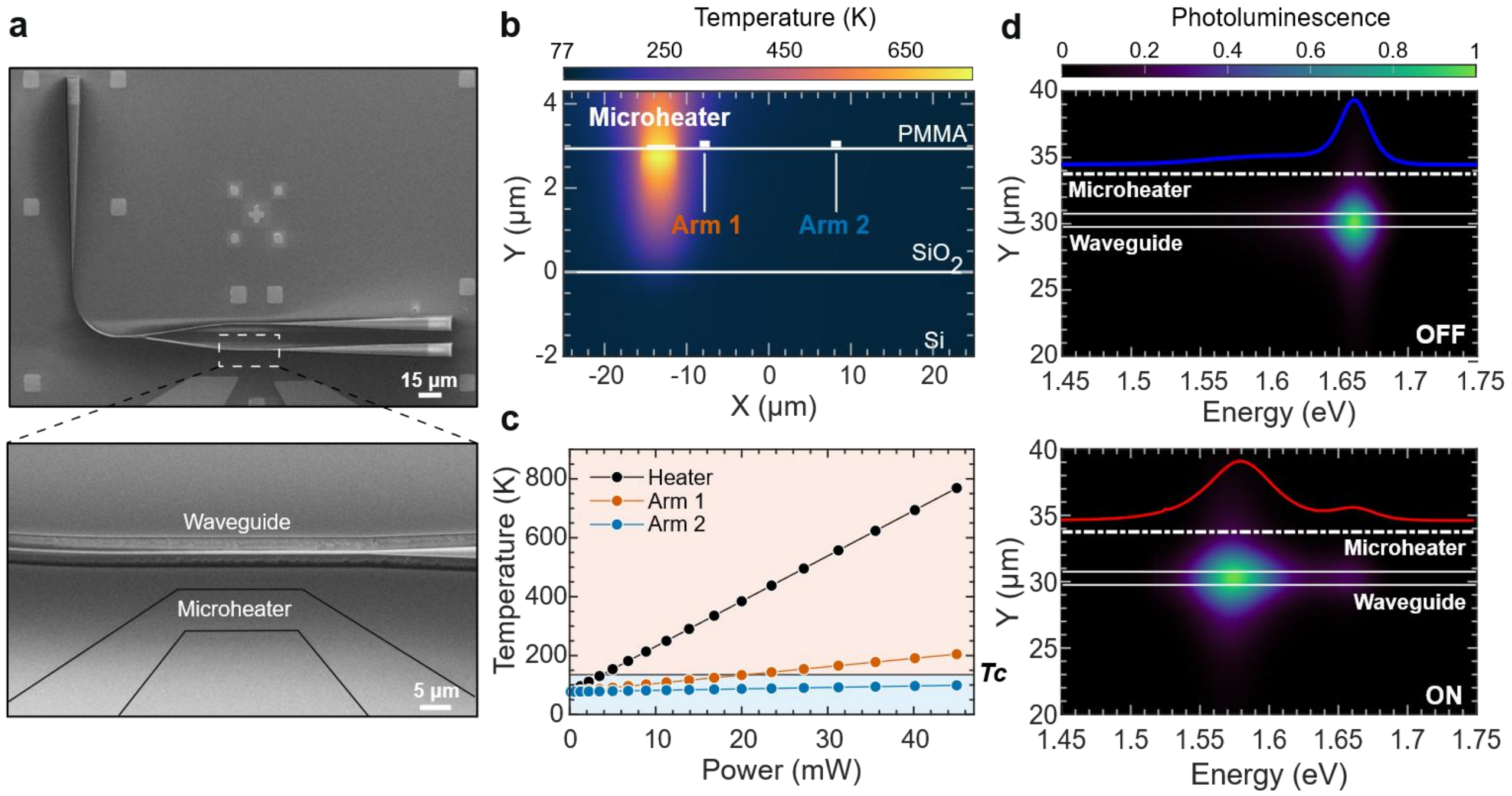

**Figure 3. Electrothermal control of the perovskite excitonic resonance: (a)** SEM image showing the polaritonic phase shifter device, that is composed by a splitter followed by the outcouplers, and the tunable electric microheater (magnified bottom panel) is placed at 3 μm distance from one of the two arms of the splitter to induce a local variation of temperature. **(b)** Numerical simulation of the temperature spatial profile showing the spread of the locally generated heat across one arm of the splitter. **(c)** Simulated temperature as a function of the applied voltage for the microheater and the two waveguide arms. Arm 1 reaches the critical phase-transition temperature $T_C$, whereas Arm 2 remains below the transition threshold over the entire power range. **(d)** Spatially resolved photoluminescence spectra measured with the heater switched off (top) and on (bottom), demonstrating an electrically induced local excitonic redshift approaching 60 meV while preserving spatial confinement to the heated waveguide arm.

As the dissipated electrical power increases, the temperature of the heated arm (Arm 1) rises above the orthorhombic-to-tetragonal transition temperature $T_C$, whereas the reference arm (Arm 2) remains below $T_c$ over the relevant operating range (**Fig. 3c**). This allows the structural transition to be selectively induced in only one arm of the device. The spatial selectivity of the perovskite thermal response was directly examined using spatially resolved photoluminescence spectroscopy (**Fig. 3d**). With the heater switched off, the excitonic emission is nearly uniform across the two arms, consistent with both regions being in the same low-temperature phase (**Fig. 3d, top**). Upon heating, the excitonic resonance redshifts only in the addressed arm, reaching a total spectral shift of approximately $60\ \mathrm{meV}$ once the local transition is completed (**Fig. 3d, bottom**). This localized excitonic reconfiguration modifies the complex refractive index and polariton dispersion along the heated section, providing the differential optical response required for phase and intensity modulation. To directly quantify the phase modulation, the real-space emission from the two output grating couplers was analysed using a Michelson interferometer comprising a mirror and a retroreflector (**Fig**. **4a**). This configuration generates two differently inverted replicas of the transmitted field, allowing the signal from Arm 1 in one interferometer path to be spatially overlapped with that from Arm 2 in the other. A small angular offset between the two replicas produces off-axis interference fringes. The fringe position is determined by the relative optical phase accumulated in the two waveguide arms, allowing the electrically induced phase shift to be extracted directly from the spatial displacement of the interference pattern. At $1.550\ \mathrm{eV}\ (800\ \mathrm{nm})$, increasing the applied voltage produces a progressive transverse displacement of the fringes, reflecting the additional phase accumulated in the heated arm (**Fig. 4b**). With the heater switched off, the interference pattern defines the initial phase difference between the two arms. As the dissipated power increases, the fringes continuously shift until, at approximately $17\text{-}20\ \mathrm{mW}$, the initial maxima approach the positions of the minima, corresponding to a phase modulation close to $\pi$. The corresponding transverse intensity profiles retain a similar spatial envelope and fringe visibility throughout the modulation (**Fig. 4c**).

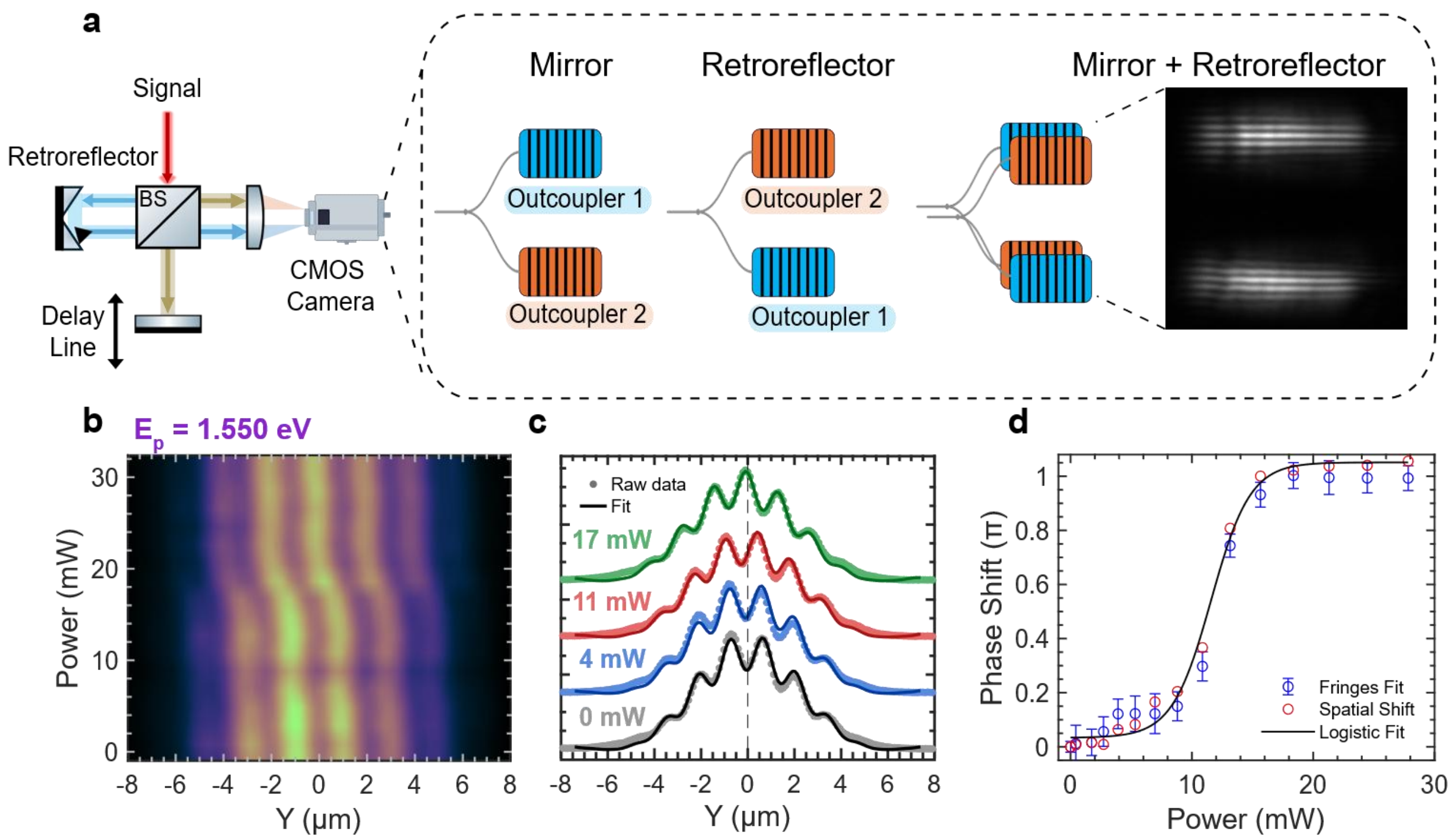

**Figure 4. Phase modulation in reconfigurable perovskite polaritonic waveguides: (a)** Schematic of the Michelson interferometer used to analyse the real-space transmission from the two output grating couplers. A mirror and a retroreflector generate differently inverted image replicas that are spatially overlapped to interfere the signals from the two output arms. The inset shows a representative experimental interference pattern. **(b)** Interference patterns recorded as a function of dissipated electrical power for the propagating polariton state at 1.550 eV (800 nm), showing a progressive spatial displacement of the fringes and the eventual inversion of maxima and minima. **(c)** Transverse intensity profiles at selected powers. The vertical dashed black line marks the position of an initial fringe maximum, which gradually evolves into a minimum as the relative phase approaches $\pi$. Circular markers represent the experimental data, while the solid lines show the corresponding fits. **(d)** Phase shift as a function of dissipated electrical power, independently extracted from fits to the interference profiles (blue circles) and from the spatial displacement of the fringe maxima (red circles). The black line shows a logistic fit to the phase values obtained from the spatial-displacement method.

The phase modulation was quantified using independent approaches. First, the interference profiles were fitted using the superposition of the two outcoupled signals, with the relative phase extracted as a fitting parameter:

$$\mathrm{I(y)} = \left(\mathrm{I_1} + \mathrm{I_2} + 2\mathcal{V}\sqrt{\mathrm{I_1 I_2}}\cos\left(\phi + \frac{2\pi}{\Lambda}\mathrm{y}\right)\right)\mathrm{sinc^2(bx)} + \mathrm{I_{bg}}$$

where, $I_1$ and $I_2$ are intensities from the outcouplers, $\mathcal{V}$ is the fringes visibility, $\Lambda$ is the fringes period, $\phi$ is the relative phase, $b$ is amplitude of the spatial envelope, and $I_{\mathrm{bg}}$ is the background intensity (**Fig. 4c**). Second, the phase variation was obtained directly from the spatial displacement of the fringe maxima relative to the zero-power condition, according to:

$$\Delta\phi = 2\pi \frac{y(P) - y(0)}{\Lambda}$$

The phase values obtained from the spatial displacement were subsequently fitted with a logistic function to describe the power-dependent evolution across the structural transition, showing a transition broadening of $\sim 8$ K (see *Supplementary Information*). The two extraction methods show close agreement over the entire investigated power range and yield a maximum phase shift of approximately $\pi$ at $1.550$ eV (**Fig. 4d**). The phase progressively increases across the transition and saturates once the exciton resonance reaches its fully shifted state, indicating completion of the orthorhombic-to-tetragonal transition within the heated section. Additional measurements confirmed the energy dependence of the phase modulation for the same modulation length. At $1.551$ eV ($795$ nm, $|X|^2 \approx 0.23$) closer to the exciton resonance, the larger excitonic fraction produces a phase shift exceeding $\pi$, whereas at $1.531$ eV ($810$ nm, $|X|^2 \approx 0.14$) the phase shift decreases to approximately $0.73\pi$ (see *Supplementary Information*). This energy dependence indicates that the interaction length required to achieve a target phase shift can be tailored according to the excitonic fraction of the propagating polariton.

**Intensity switching of propagating polaritons**

The device response changes markedly as the energy of the propagating mode is closer to the excitonic resonance. At $1.589$ eV ($780$ nm), where the propagating polariton has an excitonic fraction of $|X|^2 \approx 0.35$ both output couplers initially show comparable transmitted intensities with the heater switched off, confirming propagation through both arms of the Y-splitter (**Fig. 5a**). As electrical power is applied to the microheater, the transmitted intensity from the addressed arm progressively decreases and eventually approaches the background level, while the reference arm remains largely unaffected (**Fig. 5b**). The thermally addressed channel is therefore switched from a transmitting to a strongly attenuated state without significantly affecting propagation through the neighbouring arm. The transition between phase and intensity modulation is further quantified by comparing the extinction ratio as a function of dissipated power for different energies (**Fig. 5c**). At 1.550 eV (800 nm), where the excitonic fraction is lower, the two outputs remain nearly balanced over the full modulation range, showing that the $\pi$ phase shift is achieved with negligible change in transmitted intensity. In contrast, at 1.5895 eV (780 nm), the larger excitonic fraction results in a pronounced suppression of the addressed output once the local structural transition is activated, resulting in a modulation of depth of 1 dB/μm when the structural transition is complete.

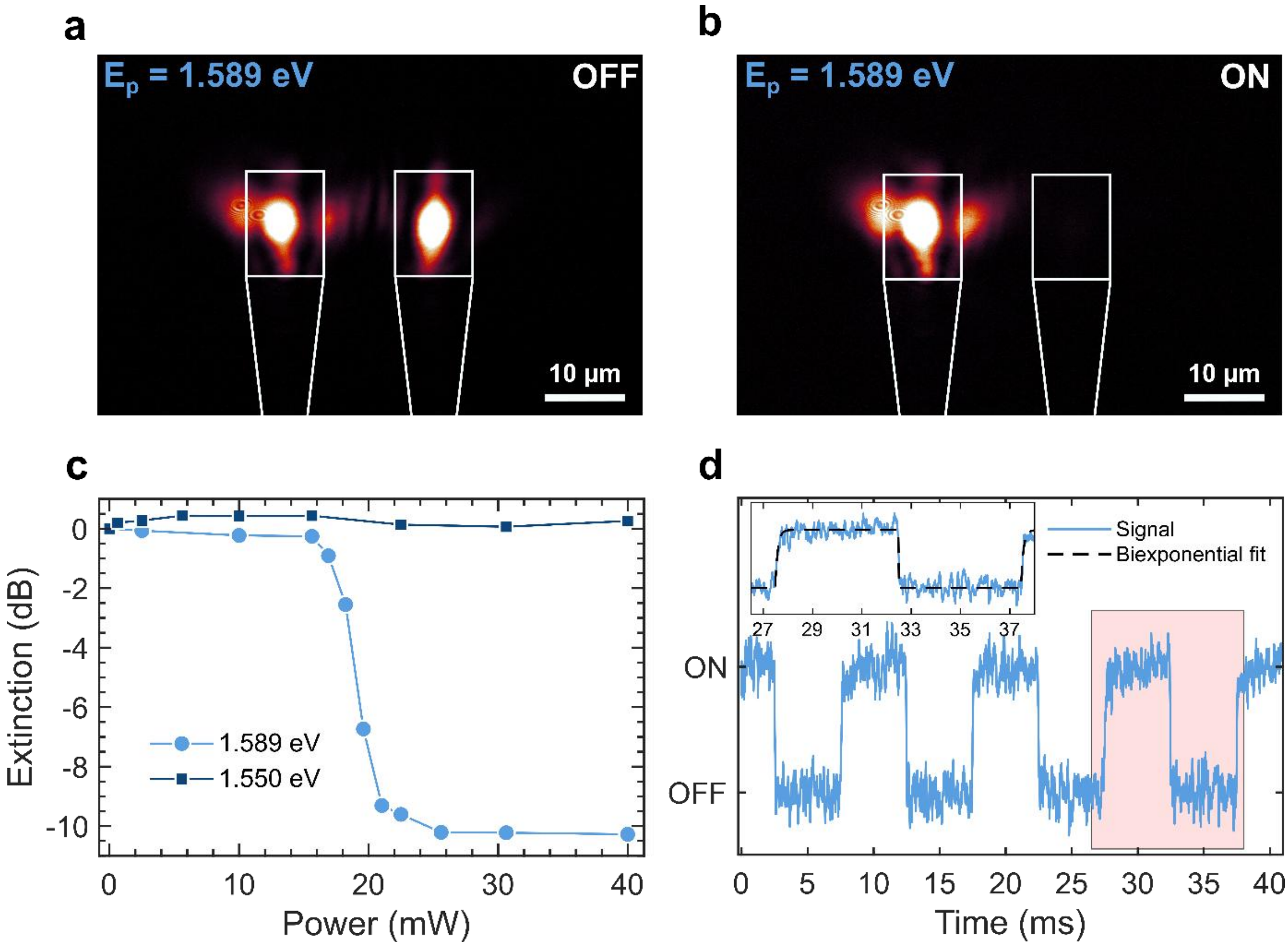

**Figure 5. Intensity modulation in perovskite polaritonic waveguides: (a-b)** Real-space emission from the two output couplers at 1.589 eV, with the heater switched off and on, respectively. Electrical heating suppresses propagation through the thermally addressed arm while preserving the output from the reference arm. **(c)** Extinction ratio between the modulated and reference outputs as a function of dissipated power. At 1.550 eV (dark blue), the two outputs remain nearly balanced, whereas at 1.5794 eV (light blue) the addressed output is progressively attenuated and ultimately switched off. **(d)** Time-resolved transmitted intensity under periodic electrical modulation of the microheater (100 Hz, 50% duty cycle), measured at 1.589 eV. The light-blue trace shows the 5-cycle phase-averaged experimental signal, and the ON and OFF levels correspond to the heater being switched on and off, respectively. Inset: magnified view of the shaded region, with the biexponential fit (black dashed line) used to extract the 10-90% ON-OFF (226 μs) and 90-10% OFF-ON (15 μs) switching times.

Measurements at intermediate energies show a correspondingly intermediate response (see *Supplementary Information*). These results demonstrate that the energy-dependent excitonic fraction governs the balance between phase and intensity modulation, allowing the same electrically reconfigurable element to operate either as a phase shifter or as an optical switch. The phase-transition dynamics was characterized through time-resolved measurements performed on a straight waveguide incorporating a 10 μm long modulated section, by recording the photon arrival times of the transmitted signal relative to the applied AC voltage (100 Hz, 50% duty cycle). The phase-averaged transmitted intensity over five modulation cycles is shown in **Fig. 5d** (see *Supplementary Information*). The OFF-ON (ON-FF) transition, defined by the 90-10 % (10-90 %) intensity change, occurs within 15 μs (226 μs), as extracted from a biexponential fit (black dashed line). These switching times correspond to an estimated 3-dB modulation bandwidth of ~23.3 kHz (~1.5 kHz), limited by the slower ON-OFF transition. The asymmetry between the on and off switching dynamics likely reflects the different thermal response of the material during heating and cooling, associated with the temperature-dependent thermal properties of the different crystalline phases of the perovskite. Comparable switching behaviour was also observed in a Mach-Zehnder interferometer with a 10 μm long modulated section, confirming that the response is preserved across different device geometries (see *Supplementary Information*). The modulation bandwidth, currently limited by heat dissipation during cooling, could be further increased through optimized heater and substrate design.

## Conclusion

We demonstrated electrically reconfigurable exciton-polariton transport in polycrystalline $MAPbI_3$ waveguide circuits with a compact active region of only 10 μm. Electrical heating drives the structural phase transition, shifting the exciton resonance by approximately 60 meV and locally reshaping the polariton dispersion with an estimated modulation bandwidth of 1.5 kHz. This enables continuous control of both phase and amplitude of the polariton mode at 20 mW, from a π phase shift to intensity modulation reaching 1 dB/μm closer to the exciton resonance. These results establish reconfiguration of an excitonic PCM as a versatile mechanism for actively controlling phase and intensity of propagating polaritons within compact integrated circuits, providing a route towards programmable polariton switches, logic elements and interferometric networks,[56,57] multiplexing, and could be extended to coupled-waveguide architectures for few-polariton nonlinear operations,[58] neuromorphic processing,[59] and synthetic-dimensional photonic lattices.[60]

## Methods

### Numerical simulations

The polaritonic circuit components were designed and optimized via Finite Difference Time Domain (FDTD) and Rigorous Coupled-Wave Analysis (RCWA) suit of the commercial software Ansys Lumerical (see *Supplementary Information*).

### Film deposition

$MAPbI_3$ films were fabricated from a 0.5 M precursor solution of $CH_3NH_3I$ (Dyesol, Greatcell) and $PbI_2$ (99.99%, TCI) with a molar ratio 1:1, in anhydrous dimethylformamide (DMF, Sigma-Aldrich). The solution was stirred for few hours at 373 K in $N_2$ inert atmosphere glovebox and then filtered by a polyvinylidene fluoride (PVDF) syringe filter (0.45 µm) before spin-coating. Prior to deposition, the substrates were cleaned with ultrasonication for 5 min in acetone and isopropanol, respectively. Subsequently, substrates were dried with nitrogen, followed by 20 minutes ozone surface treatment. The perovskite precursor solution was spin-coated onto the $Si/SiO_2$ substrates with a speed of 4000 rpm for 35 seconds using the toluene anti-solvent deposition method, with the anti-solvent being drop-cast on the substrates 6 seconds after spinning began. The film was then annealed at 373 K for 10 minutes, yielding a 140 nm thickness, as measured by atomic force microscopy (*Supplementary Information*,).

### Device fabrication

Devices were fabricated on undoped silicon wafers coated with $SiO_2$ layer (3 µm) deposited by PECVD (Applied Materials P5000). The microheaters and electrical contacts were defined in two successive laser-lithography (ML3 Pro maskless writer) and lift-off steps using AZ5214E photoresist. Chromium microheaters were first deposited by resistive thermal evaporation. In the second step, the contact electrodes were formed by sequential thermal evaporation of 3 nm Cr and 70 nm Au. Depending on the heater thickness and shape, the fabricated devcies exhibit electrical resistances ranging from approximately 0.8 to 2.3 kΩ. The $MAPbI_3$ film was subsequently deposited onto the patterned substrate. The waveguide circuits were directly structured in the perovskite layer by focused ion-beam milling (FEI Helios NanoLab 600). The in-plane waveguide structures were milled with an ion-beam current of ~6.1 pA whereas the grating couplers were patterned using a lower current of ~1.7 pA. The completed devices were encapsulated with a PMMA layer (~1.5 μm) and electrically connected by wire bonding.

**Angle-resolved spectroscopy**

Angle-resolved photoluminescence measurements were performed with a custom-built micro-spectrometer setup consisting of an inverted optical microscope (Nikon Ti-u, 50x objective, NA=0.556, 20x objective NA=0.4), a spectrograph (Andor SR-303i with a 300 lines/mm grating), and a charged-coupled detector (CCD, Andor iDus 420). A series of lenses along the beam path between the microscope and the spectrograph projects the back focal plane of the collection objective on the slit of the spectrograph, allowing the collection of angular. The sample was excited by a 400 nm continuous wave laser for the non-resonant excitation and with supercontinuum laser (Fianium) coupled to an acousto-optic modulator for the resonant excitation.

## Acknowledgements

We thank D. Gerace for his inspiring presentation on polaritonic waveguides and circuitry, which helped motivate this work. We also acknowledge the N2FC cleanroom facility at Nanyang Technological University and its staff for their technical expertise and assistance with device fabrication.

## Funding Statement

Research was supported by the Singapore Ministry of Education (MOE-T2EP50222-0015) and the National Research Foundation through the National Centre for Advanced Integrated Photonics (NRF-MSG-2023-0002).

## Author Contributions Statement

A.Z. conceived the idea and designed the experiments together with D.S. and C.S. A.Z. performed the design optimisation and numerical simulations with input from F.M. A.Z. carried out the perovskite synthesis and all fabrication steps with assistance from T.M.C, F.M., and G.A. A.Z. and M.M. built the optical setup and performed the experimental measurements. A.Z. drafted the manuscript, and all authors contributed to its review and editing. C.S. supervised the work.

## Competing Interests Statement

The authors declare no competing interests.